\documentclass[aps,prl,reprint,groupedaddress]{revtex4-1}

\usepackage{graphicx}

\usepackage{amsfonts, amsmath, amssymb,latexsym}

\usepackage{dcolumn}

\usepackage{bm}

\newcommand{\dif}{\mathrm{d}}

\begin{document}

\title{The role of attractive interactions in the dynamics of molecules 
in liquids}

\author{X. You, L. R. Pratt} 
\affiliation{Department of Chemical and Biomolecular Engineering, Tulane University, New Orleans, LA 70118}
\author{S. W. Rick} 
\affiliation{Department of Chemistry, The University of New Orleans,
New Orleans, LA 70148}

\date{\today}

\begin{abstract}The friction kernel (or memory function) $\gamma(t)$
characterizing single-molecule dynamics in strongly bound liquids exhibits two
distinct relaxations with the longer time-scale relaxation associated with
attractive intermolecular forces. This observation identifies differing roles of
repulsive and attractive interaction in the motions of molecules in equilibrium
liquids, and thus provides a basis for a renewed investigation of a van der
Waals picture of the transport properties of liquids. This conclusion is
supported by extracting $\gamma(t)$ from molecular dynamics simulation data for
four common molecular liquids.\end{abstract}

\maketitle

\section{Introduction} A basic goal of the molecular theory of liquids is the
clear discrimination of effects of intermolecular interactions of distinct
types, \emph{e.g.} excluded volume interactions and longer ranged attractive
interactions \cite{Widom:1967tz,Andersen:1976vf}. That discrimination leads to
the van~der~Waals picture \cite{Widom:1967tz,barker1976liquid,chandler1983van}
of the equilibrium theory of classical liquids. Those ideas are clear enough to
be captured in models with van der Waals \emph{limits} that are susceptible to
rigorous mathematical analysis \cite{Lebowitz:1980vo}. Ultimately, the general
theory of liquids is then founded on the composite van der Waals picture which
also serves to characterize non-van der Waals cases, such as water, for
particular scrutiny \cite{shah:144508}. Here we obtain observations that
distinguish differing roles of repulsive and attractive interaction in the
dynamics of molecules in equilibrium liquids. 

The analogue of the van der Waals picture of equilbrium liquids for transport
properties is much less developed. That is partly because of the higher variety
of transport phenomena to be addressed \cite{zwanzig1965time}. It is also
because the mathematical van der Waals limit contributes essentially at
\emph{zeroth} order to the thermodynamics, whereas extensions of van der Waals
concepts to transport parameters have shown that the leading contribution
typically comes at higher order \cite{Resibois:1972tx}. The leading contribution
from attractive interactions to a self-diffusion coefficient vanishes in the van
der Waals limit \cite{Seghers:1975ug} though the equation of state changes
qualitatively in the same limit.

\begin{center}
\begin{figure}[h]
\includegraphics[width=3.2in]{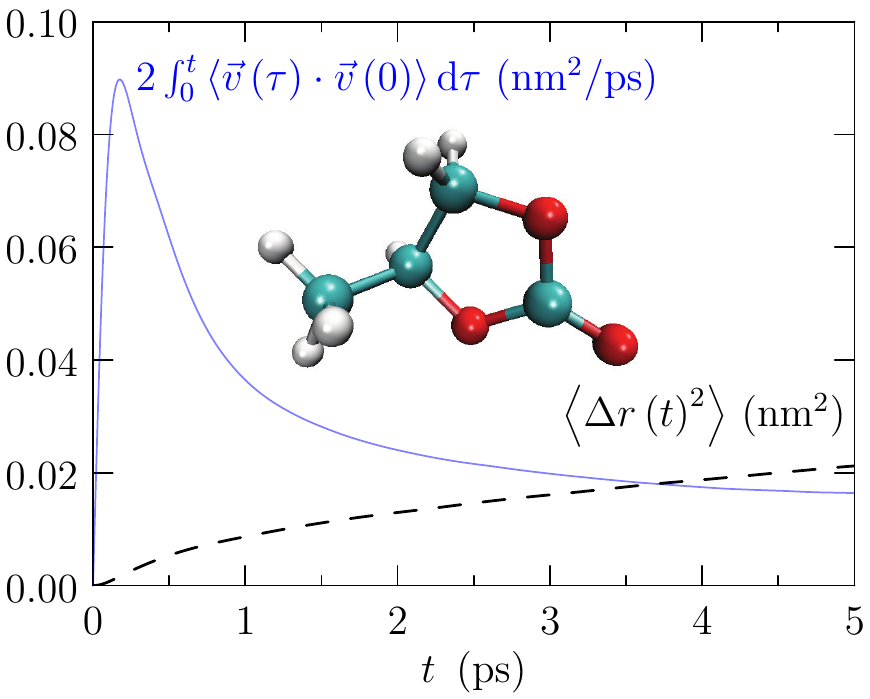}
\caption{Dashed curve: the mean-square-displacement of the center-of-mass of a
propylene carbonate molecule in liquid PC. $D_{\mathrm{PC}}=4.0\times
10^{-6}$~{cm$^2$/s}. Solid curve:  time-derivative of the
mean-square-displacement. The inset depicts the propylene carbonate (PC)
molecule.  The molecular dynamics simulation utilized  the GROMACS package in the
isothermal-isobaric ensemble (NPT) with periodic boundary conditions and $p$ = 1
atm. The GROMACS OPLS all-atom force field was adopted for liquid PC, and
temperature $T$ = 300K maintained by Nose-Hoover thermostat. The system of $n=
1000$ PC molecules was aged for 10~ns, then a 1~ns trajectory was obtained,
saving configurations every 10th 1 fs time step for analysis. }
\label{fig:msdPC}
\end{figure}
\end{center}

Nevertheless, it is important accurately to characterize the contributions of
realistic attractive interactions to kinetic characteristics of liquids. This
has been the serious topic of previous investigations
\cite{Verlet:1967cm,Kushick:1973ir}. One distinction of the work here from
previous efforts is that we focus  on a specific autocorrelation function
of the random forces on a molecule in the liquid, $\gamma(t)$ defined below.
Another distinction is that we consider practical examples of solvent liquids
that are strongly bound compared to the Lennard-Jones (LJ) models that have been
the focus of historical work. We characterize this `strongly bound' distinction
by the ratio of the critical-point to the triple-point temperatures
($T_\mathrm{c}/T_\mathrm{t}$). For the LJ fluid this ratio is
$T_\mathrm{c}/T_\mathrm{t}=1.9$, but here we consider propylene carbonate (PC:
3.5), ethylene carbonate (EC: 2.3), acetonitrile (AN: 2.4), and water (W: 2.4).
In these practical cases, attractive interactions leading to the greater
binding strength are more prominent than in the historical LJ work.

\begin{center}
\begin{figure}[h]
\includegraphics[width=3.2in]{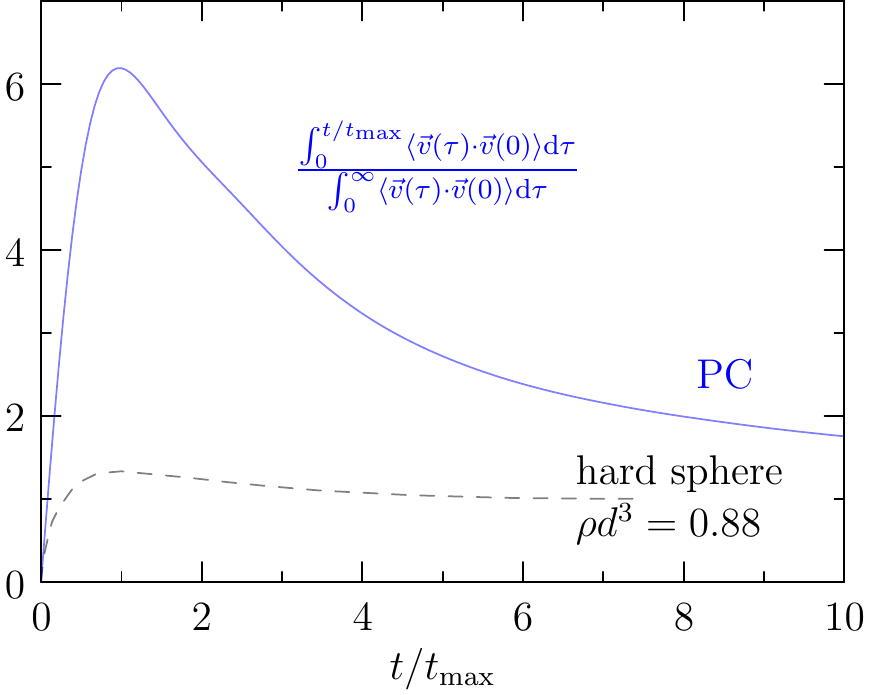}
\caption{Normalized time-derivative of the mean-square displacement of a
molecule center-of-mass. The blue curve is redrawn from FIG.~\ref{fig:msdPC},
with the vertical scaling so that  the graph equals one (1) at
large time, and the horizontal scaling so that the graph has a maximum
at the time $t_\mathrm{max}.$ The dashed curve is the result for the hard
sphere fluid, redrawn from Alder, \emph{et al.} \cite{Alder:1970fg} The density,
$\rho d^3 = 0.88,$ is within about 5\% of the hard-sphere freezing point. The
maximum in these graphs occurs when $C(t)$ changes sign. The result for the
realistic PC fluid is qualitatively different from the result for the hard
sphere case.}
\label{fig:AWcompare}
\end{figure}
\end{center}

Primitive results that motivate our observations are shown in
FIGs.~\ref{fig:msdPC} and \ref{fig:AWcompare}. The mean-square-displacement
$\left\langle\Delta r(t)^2\right\rangle$ of the center-of-mass of a PC molecule
in liquid PC (FIG.~\ref{fig:msdPC}) achieves growth that is linear-in-time only
after tens of collision-times. The time-derivative of $\left\langle\Delta
r(t)^2\right\rangle$ displays a prominent maximum that can be taken as a rough
marker for a collision time. After that, the derivative decreases by more than a
factor of ten, reflecting the integrated strength of a negative tail of the
velocity autocorrelation function, 
\begin{eqnarray}
C(t) = \left\langle \vec{v}\left(t\right)\cdot 
\vec{v}\left(0\right)\right\rangle/\left\langle v^2\right\rangle~.
\end{eqnarray}
That negative tail is qualitatively different from the case of the hard-sphere
fluid at high densities (FIG.~\ref{fig:AWcompare}) \cite{Alder:1970fg}. 
More recent studies of atomic fluids with purely repulsive inter-atomic forces
show that negative tails in the velocity autocorrelation
functions are slight \cite{HEYES:1998eq,Heyes:2002gu}. The
result for the realistic PC fluid is qualitatively different from the atomic
repulsive force case.

\section{Methods}
We focus on the friction kernel (memory function), $\gamma\left(t\right)$ 
defined by
\begin{eqnarray}
M\frac{\dif C(t)}{\dif t} = - \int_0^t \gamma\left(t-\tau\right)C\left(\tau\right) 
\dif\tau~,
\label{eq:gle}
\end{eqnarray}
where $M$ is the mass of the molecule. $\gamma\left(t\right)$ can be considered
the autocorrelation function for the \emph{random} forces on a
molecule \cite{forster1975hydrodynamic}. The textbook method for extracting
$\gamma\left(t\right)$ utilizes standard Laplace transforms. But inverting the
Laplace transform is non-trivial and we have found the well-known Stehfest
algorithm \cite{Stehfest:1970vj} to be problematic.   Berne and Harp \cite{berne1970calculation}
developed a finite-difference-in-time procedure for extracting $\gamma\left(t\right)$
from Eq.~\eqref{eq:gle}.  That procedure is satisfactory but sensitive 
to time resolution in the numerical $C\left(t\right)$ that is used as input 
here.  Another alternative expresses the Laplace transform as 
Fourier integrals, utilizing specifically the transforms
\begin{subequations}
\label{eq:ft}
\begin{align}
\hat{C}'\left(\omega\right) & =  \int_0^\infty C(t) \cos \left(\omega t\right) \dif t~, \\
\hat{C}''\left(\omega\right) & =  \int_0^\infty C(t) \sin \left(\omega t\right) \dif t ~.
\end{align}
\end{subequations}
Then
\begin{eqnarray}
\int_0^\infty \gamma(t) \cos \left(\omega t\right) \dif t = 
\frac{M \hat{C}'\left(\omega\right) }{\hat{C}'\left(\omega\right)^2 + 
\hat{C}''\left(\omega\right)^2}~.
\end{eqnarray}
Taking $\gamma\left(t\right)$ to be even time, the cosine transform is
straightforwardly inverted. $\Omega^2 = \left\langle
F^2\right \rangle/3Mk_{\mathrm{B}}T,$ with  $F=\vert
\vec{F}\vert$ the force on the molecule, provides the normalization
$\gamma(0) = M\Omega^2.$

\section{Results and Discussion} 
\begin{center}
\begin{figure}[h]
\includegraphics[width=3.2in]{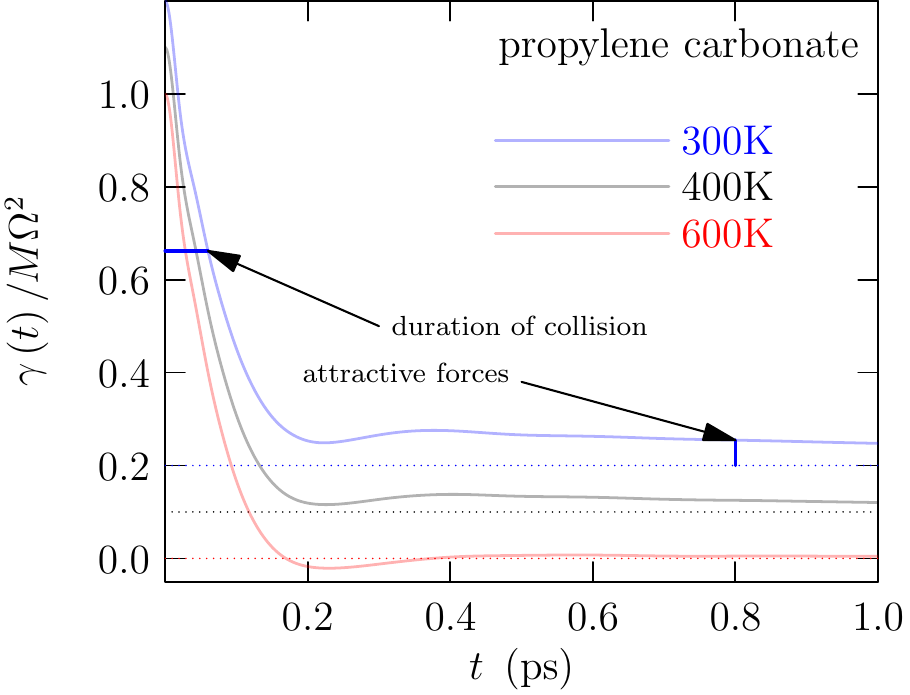}
\caption{Autocorrelation function for the random forces on the center-of-mass
of a propylene carbonate molecule as a function of temperature at 
constant pressure, $p=1$~atm.  The longer time-scale relaxation becomes 
less prominent at higher $T$.}
\label{fig:GammabyMtPC}
\end{figure}
\end{center}

The two numerical methods for extracting
$\gamma(r)$ from $C(t)$ agree well (FIG.~\ref{fig:array}).
$\gamma\left(t\right)/M\Omega^2$ for four strongly bound liquids are
qualitatively similar to each other and show two distinct relaxations. The
historical LJ results \cite{Verlet:1967cm,Kushick:1973ir} are consistent with
this, though the two relaxation behaviors are distinct for the LJ fluid only at
the lowest liquid temperatures \cite{Kushick:1973ir}. We suggest that the
slowest-time relaxation derives from molecularly long-ranged interactions, while
the fastest-time relaxation is associated with collisional events and
short-ranged interactions. This discrimination of long-ranged and short-ranged
interaction effects was expressed by Wolynes \cite{Wolynes:1978kc} long-ago in
the context of ion mobilities in solution. Here mobilities of non-ionic species
are considered, though similar behavior has been identified in just the same way
for organic ions in solution \cite{zhu2012pairing}. The results for liquid PC at
several higher temperatures (FIG.~\ref{fig:GammabyMtPC}) show that the amplitude
of this longer-time-scale decay decreases with increasing $T$, as expected.

\begin{widetext}
\begin{center}
\begin{figure}[h]
\includegraphics[width=5.0in]{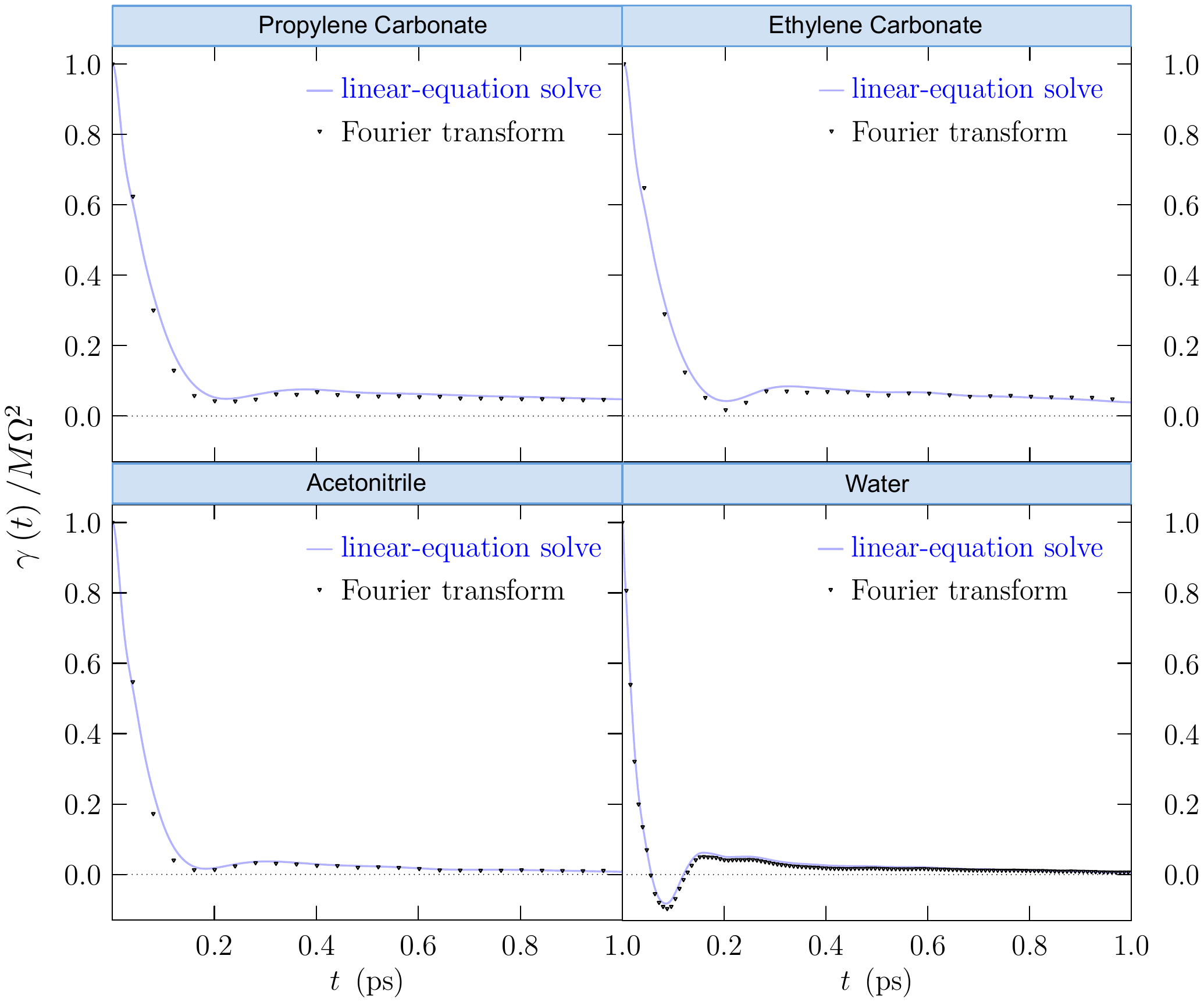}
\caption{ $\gamma(t),$ the friction kernel (or memory function). All the
simulations used the GROMACS package and periodic boundary conditions,
with  $p$ = 1 atm in the isothermal-isobaric ensemble. PC calculations were
specified with FIG.~\ref{fig:msdPC}. For the acetonitrile, simulation the
force-field of Nikitin and Lyubartsev \cite{nikitin2007new} was assumed. For
ethylene carbonate, we used the GAFF force field \cite{gaff}, a system size of
215 molecules at $T=313$ K. For the water simulation, the TIP4P-EW model
\cite{Horn:2004hi} was used, and the trajectory was saved every 1 fs. } \label{fig:array}
\end{figure}
\end{center}
\end{widetext}

\begin{center}
\begin{figure}[h]
\includegraphics[width=3.2in]{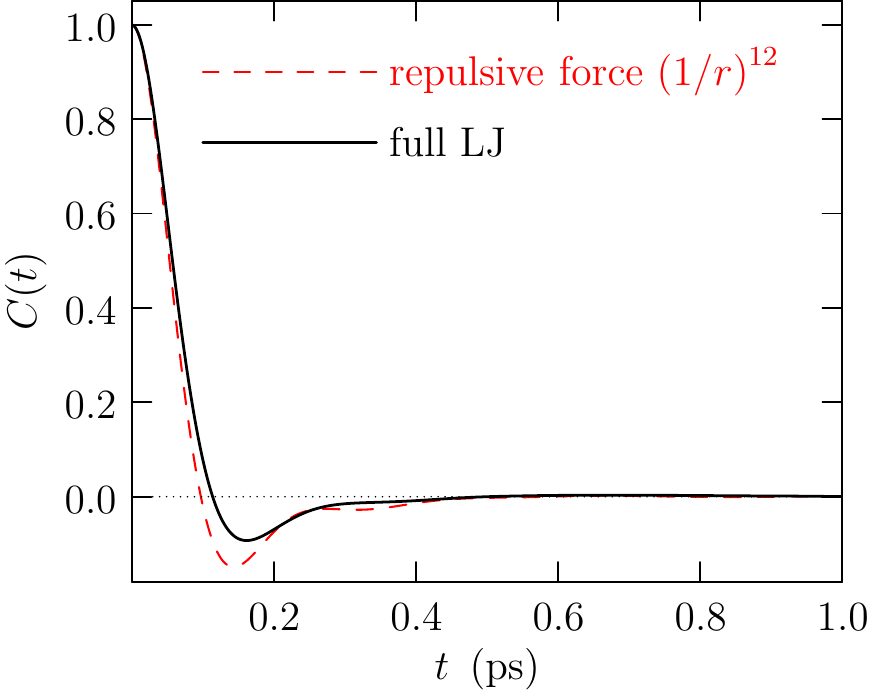}
\caption{Velocity autocorrelation function for the LJ 6-12 fluid and the
corresponding result when the $1/r^6$ contribution to the pair potential energy
is dropped. The LJ thermodynamic state point is $\rho \sigma ^3 = 1.06,$ and
$k_\mathrm{B}T/\epsilon = 3.66 \approx 2.8 T_\mathrm{c}$ \cite{Smit:1992hp}. For
the $1/r^{12}$ case, this density is slightly higher than a high density cases
studied by Heyes \emph{et al.} \cite{Heyes:2002gu}, (effective packing fraction
$\xi_{\mathrm{HS}} \approx 0.466$ compared to 0.45). } \label{fig:vacfCompare}
\end{figure}
\end{center}

By comparison, Dang and Annapureddy \cite{Dang:2013fb,Annapureddy:2014isa}
evaluated $\gamma(t)$ for Dang-Chang-model water \cite{Dang:1997gv} in a
different setting, and they obtained the distinct bi-relaxation
observed here. In that alternative setting, the separation between a water
oxygen atom and a near-neighbor ion was constrained at a barrier value and $t>0$
negative values of $\gamma(t)$ (FIG.~\ref{fig:array}) were not observed
\cite{Dang:2013fb,Annapureddy:2014isa}.

\begin{center}
\begin{figure}[h]
\includegraphics[width=3.2in]{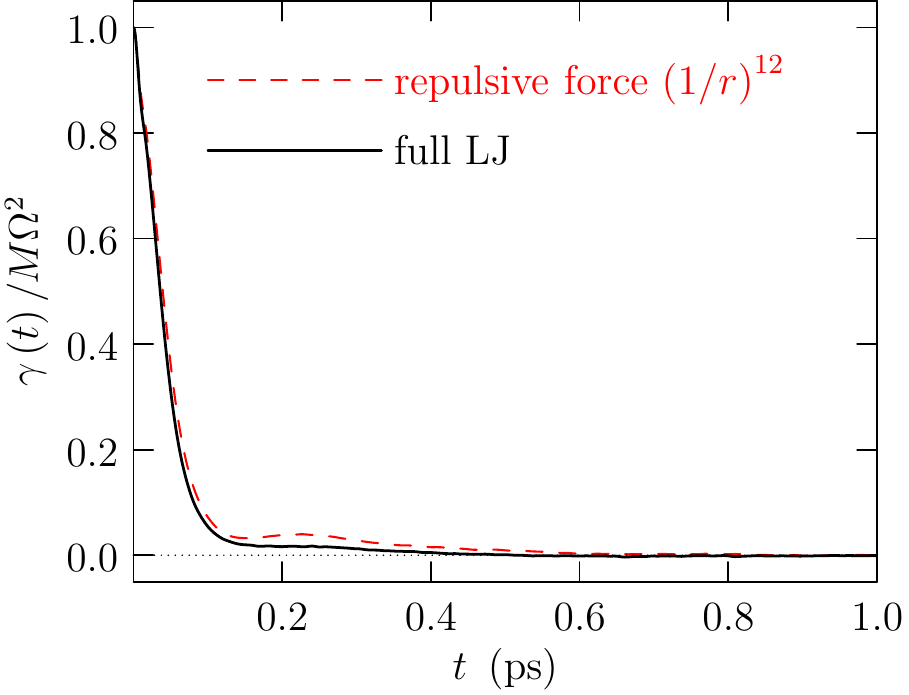}
\caption{For the LJ system of FIG.~\ref{fig:vacfCompare}. The second,
longer-time-scale relaxation is not evident at this thermodynamic state which is
a high density state, super-critical  for the LJ liquid. Thus, the
longer-time-scale relaxation of Fig.~\ref{fig:array} for water, is due to the
electrostatic interactions that contribute largely to the cohesive binding of
liquid water.} 
\label{fig:Compare}
\end{figure}
\end{center}

The longer-time scale decay of $\gamma(t)$ (FIG.~\ref{fig:array}) is less
prominent for the liquid water case than for the other cases. We investigated
this further by eliminating the partial charges associated with the
pair-molecule interactions, leaving LJ 6-12 interactions
(FIGs.~\ref{fig:vacfCompare} and \ref{fig:Compare}). That underlying LJ case is
strongly super-critical. Furthermore, with the implied high-density of the
underlying LJ system, the continuous-repulsive-force $\left(1/r^{12}\right)$
case in not similar to the result from the previous study of Heyes \emph{et al.}
\cite{Heyes:2002gu}; we observe a distinct recoil feature at this density.

\section{Conclusions} For strongly bound liquids, the friction kernel (or memory
function) $\gamma(t)$ (Eq.~\eqref{eq:gle}) exhibits two distinct relaxations
with the longer time-scale relaxation associated with attractive intermolecular
forces.

\section{Acknowledgement} 

This work was supported by the National Science Foundation under the NSF EPSCoR
Cooperative Agreement No. EPS-1003897, with additional support from the
Louisiana Board of Regents.


%

\end{document}